# Sustainable Transformer Neural Network Acceleration with Stochastic Photonic Computing

S. Afifi[1], O. Alo[2], I. Thakkar[2], and S. Pasricha[1]

1. Department of Electrical and Computer Engineering, Colorado State University, Fort Collins, CO, USA
2. Department of Electrical and Computer Engineering, University of Kentucky, Lexington, KY, USA
*sudeep@colostate.edu*

*Abstract*—Transformers achieve state-of-the-art performance in natural language processing, vision, and scientific computing, but demand high computation and memory. To address these challenges, we present ASTRA, the first silicon-photonic accelerator leveraging stochastic computing for transformers. ASTRA employs novel optical stochastic multipliers and unary/analog homodyne accumulation in a crosstalk-minimal organization to efficiently process dynamic tensor computations. Evaluations show at least 7.6× speedup and 1.3× lower energy overheads compared to state-of-the-art accelerators, highlighting ASTRA's potential for efficient, scalable transformer inference.

## I. Introduction

Transformers have become central to natural language processing, computer vision, and scientific computing, but their billions of parameters demand massive computation and memory, straining conventional CPUs, GPUs, and TPUs [1]-[2]. As transistor scaling slows, these platforms struggle to sustain energy efficiency. Photonic computing offers high bandwidth, parallelism, and low latency, yet existing accelerators face challenges including insertion loss, heterodyne crosstalk, reliance on DACs, and poor support for the dynamic dataflows of transformers [3]. In parallel, stochastic computing has emerged as a promising paradigm that simplifies multiplications into bitwise operations, offering lower circuit complexity and improved energy efficiency, though it traditionally faces accuracy limitations [4]. This work introduces ASTRA, the first optical accelerator to leverage stochastic computing for transformer neural network acceleration.

## II. ASTRA Hardware Accelerator

ASTRA introduces a new paradigm for efficient transformer inference by combining silicon photonics with stochastic computing. Unlike prior photonic accelerators that rely on multi-level analog amplitude encoding and weight-stationary mappings, ASTRA shifts optical computation to a binary-temporal paradigm. Its novel vector dot-product (VDP) core integrates hundreds of optical stochastic signed multipliers (OSSMs) to accelerate transformer operations. Each OSSM (Fig. 1) performs multiplications as bitwise AND operations using optical AND gates (OAGs; Fig. 2), eliminating DACs and reducing power. In this design, multiplication complexity is moved from optical amplitude precision to stochastic temporal density, simplifying modulation and scaling. Furthermore, ASTRA unifies stochastic-to-binary conversion, accumulation, and optical-to-electrical transduction within compute-capable transducer units, reducing peripheral overheads and improving scalability (Fig. 3). Accuracy is preserved through unary/analog-domain accumulation, while homodyne VDPEs with up to 1024 OSSMs remove crosstalk [6], minimize insertion losses, and enable massive parallelism. By dynamically encoding operands in the optical domain, ASTRA enables flexible output-stationary dataflows, reducing reconfiguration time and data movement.

Our key contributions include: 1) the first stochastic-photonic accelerator tailored for transformers, 2) novel OSSMs enabling signed, full-range multiplications, and 3) architecture-level optimizations to reduce data movement and conversion overheads. Evaluation results showing up to 7.6× speedup and 1.3× energy reduction over state-of-the-art accelerators, with >1000× energy savings versus CPUs, GPUs, and TPUs. By reducing conversion costs, minimizing insertion losses, and exploiting photonic parallelism, ASTRA delivers a scalable and sustainable platform for transformer acceleration.

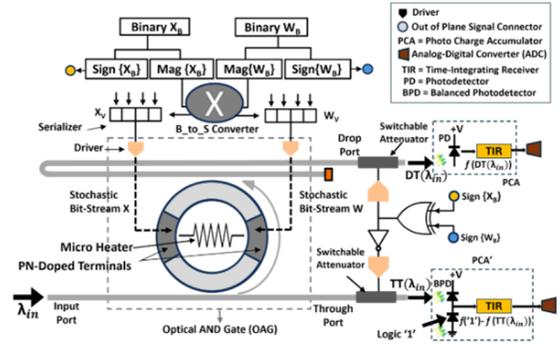

Fig. 1. Schematic of our Optical Stochastic Signed Multiplier (OSSM) [5].

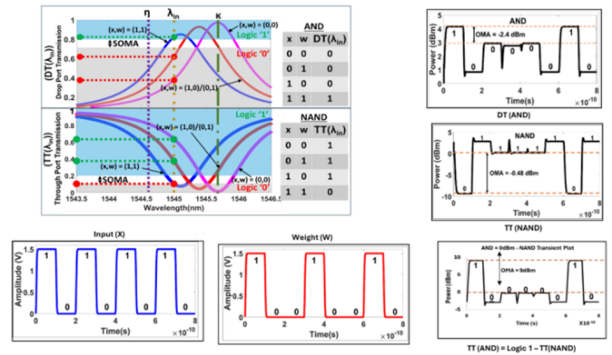

Fig. 2. (a) Operation of optical AND gate (OAG), (b) input X and weight W bit streams used for analysis, (c) results of OAG's transient analysis [5].

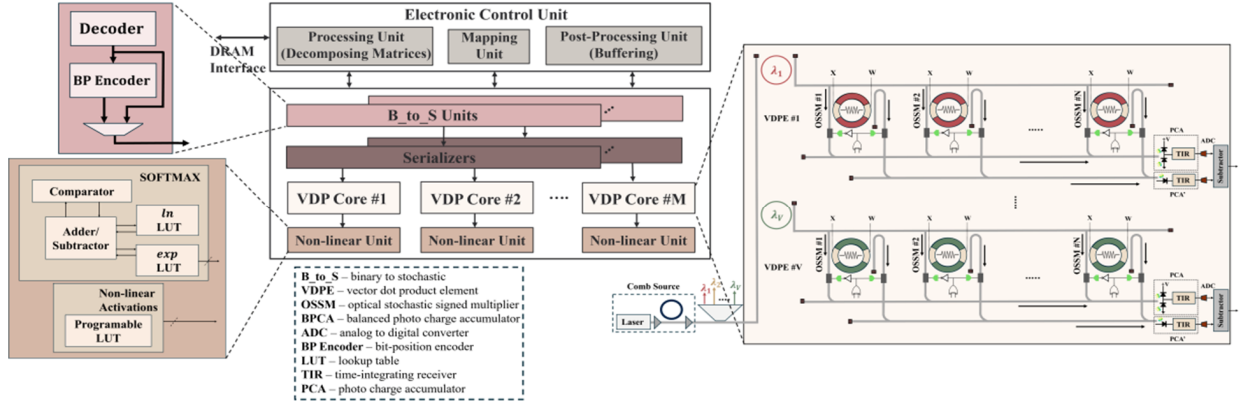

Fig. 3. ASTRA architecture overview showing vector dot-product (VDP) cores, non-linear units, binary-to-stochastic (B-to-S) circuits, and serializers [5].

## III. EVALUATION

We evaluated ASTRA through device- and architecture-level simulations on five transformer models: Transformer-base, BERT-base, Albert-base, ViT-base, and OPT-350. A custom simulator modeled layer mapping, peripheral devices, and photonic effects such as propagation, splitter, and resonator losses, while electronic components were characterized with CACTI and Vivado; models were trained and evaluated in PyTorch. Using 8-bit quantization with 128-bit stochastic streams plus a sign bit preserved accuracy within 1.2% of FP32 across NLP, vision, and generation tasks.

ASTRA achieves large-scale parallelism by supporting over 1,000 OAGs per wavelength at >30 Gbps speeds, enabled by low-power homodyne VDPEs and efficient photo-charge accumulators. Detailed device-level analysis shows that each OAG operates at approximately 0.5μW optical power after accounting for insertion and propagation losses, enabling up to 1024 OAGs per wavelength without increasing per-wavelength laser power requirements [7]. This scalability is achieved by reducing optical dynamic range via binary ON/OFF stochastic encoding rather than multi-level amplitude modulation. Further, temporal analog accumulation in compute-capable transducer units allows in-situ summation of stochastic multiplication results, avoiding costly reductions and stochastic additions.

This design provides massive throughput while operating at low optical power levels, ensuring scalability without excessive energy costs (Fig. 4). Component-wise analysis (Fig. 5) shows that serializers and OAGs dominate energy usage due to transformer matrix sizes, yet overall efficiency is improved by eliminating DACs, limiting ADC use to final outputs, and performing in-situ accumulation. Compared with CPUs, GPUs, TPUs, and other accelerators (FPGA_ACC, TransPIM, LT, TRON, and SCONNA), ASTRA consistently demonstrates superior energy efficiency (Fig. 6). Normalized to CPU, it achieves at least 1.3× lower energy costs, outperforming even photonic baselines by using lightweight OSSMs, homodyne VDPEs, and optimized dataflows. These results validate that ASTRA's binary-temporal photonic architecture enables scalable, low-dynamic-range optical computation while maintaining high throughput and energy efficiency.

## ACKNOWLEDGEMENTS

This research was supported in part by National Science Foundation grant CCF-2450615.

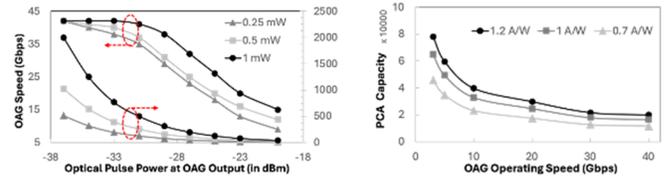

Fig. 4. Vector dot product engine (VDPE) scalability results.

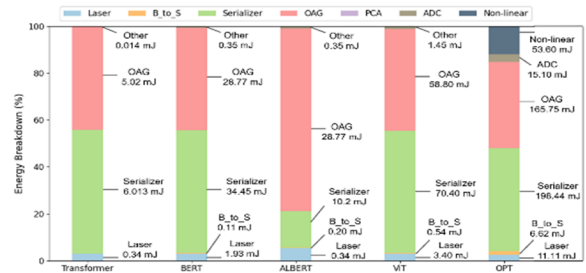

Fig. 5. Energy breakdown across ASTRA components.

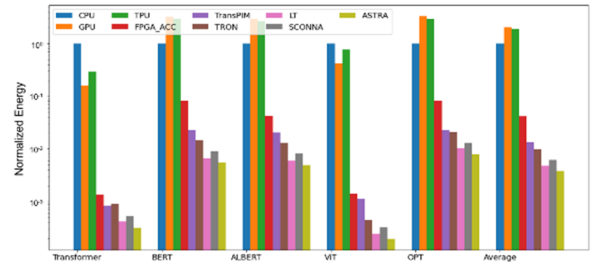

Fig. 6. Energy comparison results.